\documentclass[12pt,preprint]{aastex}
\usepackage{epsf}

\begin{document}
\newcommand{\msun}{M_{\odot}}
\newcommand{\kms}{\, {\rm km\, s}^{-1}}
\newcommand{\cm}{\, {\rm cm}}
\newcommand{\gm}{\, {\rm g}}
\newcommand{\erg}{\, {\rm erg}}
\newcommand{\kel}{\, {\rm K}}
\newcommand{\kpc}{\, {\rm kpc}}
\newcommand{\mpc}{\, {\rm Mpc}}
\newcommand{\seg}{\, {\rm s}}
\newcommand{\kev}{\, {\rm keV}}
\newcommand{\hz}{\, {\rm Hz}}
\newcommand{\etal}{et al.\ }
\newcommand{\yr}{\, {\rm yr}}
\newcommand{\gyr}{\, {\rm Gyr}}
\newcommand{\eq}{eq.\ }
\def\arcsec{''\hskip-3pt .}

\title{
Star Captures by Quasar Accretion Disks: A Possible Explanation of the
$M-\sigma$ Relation }
\author{ Jordi Miralda-Escud\'e$^{1,2}$ \& Juna A. Kollmeier$^1$}
\affil{${}^1$ The Ohio State University}
\affil{${}^2$ Institut d'Estudis Espacials de Catalunya/ICREA}
\email{jordi@astronomy.ohio-state.edu, jak@astronomy.ohio-state.edu}

\begin{abstract}
A new theory of quasars is presented in which the matter of thin
accretion disks around black holes is supplied by stars that plunge
through the disk. Stars in the central part of the host galaxy are
randomly perturbed to highly radial orbits, and as they repeatedly cross
the disk they lose orbital energy by drag, eventually merging into the
disk. Requiring the rate of stellar mass capture to equal the mass
accretion rate into the black hole, a relation between the black hole
mass and the stellar velocity dispersion is predicted of the form
$M_{BH}\propto \sigma_*^{30/7}$. The normalization depends on various
uncertain parameters such as the disk viscosity, but is consistent with
observation for reasonable assumptions. We show that a seed central
black hole in a newly formed stellar system can grow at the Eddington
rate up to this predicted mass via stellar captures by the accretion
disk. Once this mass is reached, star captures are insufficient to
maintain an Eddington accretion rate, and the quasar may naturally turn
off as the accretion switches to a low-efficiency advection mode. The
model provides a mechanism to deliver mass to the accretion disk at
small radius, probably solving the problem of gravitational instability
to star formation in the disk at large radius. We note that the matter
from stars that is incorporated to the disk has an average specific
angular momentum that is very small or opposite to that of the disk,
and discuss how a rotating disk may be maintained as it captures this
matter if a small fraction of the accreted mass comes from stellar
winds that form a disk extending to larger radius. We propose several
observational tests and consequences of this theory.
\end{abstract}


\section{Introduction}

  The basic model for how quasars are able to emit their prodigious
radiative luminosities from a very small region of space has been in
place for a long time (Lynden-Bell 1969): a massive black hole in the
center of a galaxy accretes from a thin gaseous disk, converting $\sim$
10\% of the rest-mass of the gas into the radiation that is emitted. The
gas in the disk is heated by viscous processes as it accretes, providing
energy for radiating the continuum optical-ultraviolet emission from the
hot, optically thick surface of the disk (Shakura \& Sunyaev 1973;
Pringle 1981).

  Black holes have now been detected in the centers of many galaxies
and found to correlate strongly with the presence of a spheroidal
stellar component, either the bulge of a spiral galaxy or an elliptical
galaxy (e.g., Kormendy \& Richstone 1995; Richstone \etal 1998,
Magorrian \etal 1998). The mean mass
density in the universe of nuclear black holes (i.e., black holes found
in galactic nuclei and presumed to be responsible for galactic nuclear
activity such as the quasar phenomenon), $\rho_{BH}$, should be related
to the integrated emission from all active galactic nuclei over the past
history of the universe according to
\begin{equation} 
{\epsilon \over 1-\epsilon} \, \rho_{BH} c^2 =
\int e(z)\, (1+z)\, dz ~,
\end{equation}
where $e(z)\, dz$ is the present energy density
in radiation coming from AGN in the redshift range $z$ to $z+dz$, and
$\epsilon$ is the mean radiative efficiency of accretion (Soltan 1982).
This relation is consistent with present observations with $\epsilon
\simeq 0.1$ (e.g., Barger \etal 2001; Aller \& Richstone 2002; Yu \&
Tremaine 2002; Haehnelt 2003), implying that accretion of gas from thin
disks likely played the dominant role in the growth of nuclear black
holes. In addition, observations have shown that the black hole mass,
$M_{BH}$, is tightly related to the stellar velocity dispersion,
$\sigma_*$, as $M_{BH} \propto \sigma_*^a$, where $a$ is in the range
$4$ to $4.5$ (Gebhardt \etal 2000; Merritt \& Ferrarese 2000, 2001;
Tremaine \etal 2002). This relation suggests that there is some
connection between the accretion activity of the black hole (which
determines its final mass) and the stellar system that surrounds it.

  The formation and growth of nuclear black holes from an accretion disk
poses a number of outstanding problems: how is the large amount of mass
that must be fed to the black hole funneled from the typical sizes of
galaxy spheroids ($\sim 1$ kpc) into the tiny region in the center of a
galaxy where the mass is dominated by the central black hole ($\sim 1$
pc), and into the inner accretion disk where most of the energy of
active galactic nuclei is radiated ($\sim 10^{-2}$ to $10^{-5}$ pc) ?
What prevents this gas from turning into stars well before coming close
to the central accretion disk, as normally occurs in galactic gaseous
disks and in galaxies with an irregular distribution of gas? Once the
gas is in the accretion disk, what happens when the disk becomes
self-gravitating in its outer parts (e.g., Shlosman, \& Begelman 1987;
Goodman \& Tan 2003), and hence unstable to form stars? Why is the black
hole mass tightly related to the velocity dispersion of the stellar
system around it, when the physical scales of these components are so
vastly different?

  Another possibility for the growth of black holes is by the direct
capture of stars that are randomly perturbed into high-eccentricity
orbits. Zhao, Haehnelt, \& Rees (2002) found that this mechanism leads
to a relation $M_{BH} \simeq 10^8 \msun (\sigma_{\*}/ 200 \kms)^5
(t_0/14 \gyr)$, where $t_0$ is the total time during which stars can be
captured by the black hole. This is valid if one makes the approximation
of a full loss-cylinder
\footnote{We use the term loss-cylinder to refer to the region of
phase-space from which stars will be captured. At a fixed point in
space in a spherical potential, the shape of this region in velocity
space is a narrow cylinder along the radial velocity axis. This has
usually been referred to as loss-cone in the literature.},
i.e., one assumes that when stars are captured their orbits are
immediately replenished by relaxation processes.
The slope of this relation is
in good agreement with the observations (see also Merritt \& Poon 2003).
However, in this model black holes do not grow by gas accretion,
so the similarity of the black
hole mass density and the energy density from quasars at present is not
accounted for. In addition, black holes would grow over a timescale too
long to account for their presence at the high redshifts at which
quasars are observed. Moreover, the relation $M_{BH}\propto
\sigma_*^{5}$ holds only for black hole masses $M_{BH} \gtrsim 10^8
\msun$, which can swallow normal stars whole without tidally disrupting
them, as discussed by Zhao \etal (2002).

  We propose a different idea in this paper. Black holes grow during the
quasar epoch by accreting gas from a thin accretion disk in the standard
way. At the same time, stars from the stellar system around the black
hole are captured into the accretion disk when their orbits become
highly eccentric and they plunge through the disk. Even though the stars
are slowed down by only a small fraction of their velocity due to the
drag force at every disk crossing, repeated crossings result in their
final merger into the disk. The capture of stars by an accretion disk,
discussed by Ostriker (1983), Syer, Clarke, \& Rees (1991), Artymowicz
\etal (1993), and Zurek \etal (1994), can have a much higher cross
section than direct capture by the black hole, shortening the time
required for the black hole to grow. The growth of black holes from a
gaseous disk that is continuously replenished with matter from plunging
stars can also solve the problem of how matter is transported from the
galactic system to the very small central accretion disk, and provides
a way to connect the final mass of the black hole after accretion stops
with the velocity dispersion of the surrounding stellar system.

  Delivering mass to the accretion disk by means of stars that are
randomly scattered into the loss-cylinder implies that the structure of
the disk should change in an important way owing to the addition of
energy and angular momentum to the disk as the plunging stars dissipate
their kinetic energy during disk crossings and the angular momentum of
the stars is incorporated to the disk. The issue of the global angular
momentum will be discussed in \S 5. In this paper, we will generally
assume a steady-state disk structure ignoring the effects of the stars,
leaving for later work a fully self-consistent model in which the
effect of the plunging stars added to the disk is taken into account.

  The model is presented in detail in \S 2, where the condition for
stars to be captured by the disk is described and a resulting
$M_{BH}-\sigma$ relation is inferred. In \S 3 we discuss how the
generic problem of self-gravity of the disk may be solved in our model,
and in \S 4 we comment on the fate of the stars after they are embedded
in the disk. \S 5 describes the total disk angular momentum problem and
proposes a solution. Finally, \S 6 discusses the predictions of the
model and presents the conclusions.

\section{Black Hole Growth from Collisions of Stars with the Accretion Disk}

  Before going into the detailed description of our model, it will be
useful to give an overview of our goals in this section. We will start
by reviewing steady-state accretion disk models, deriving the surface
density profile. We will then infer the condition required for a
typical main-sequence star to be captured by the disk as it slows down
in multiple disk crossings. As we shall show below, captured stars are
eventually destroyed and their matter is dispersed within the disk. This
will lead us to a rate at which mass is being delivered to the disk by
the plunging stars. We will then require that this mass delivery rate
from stars into the disk is equal to the mass accretion rate of the disk
gas into the black hole, which is given in terms of the quasar
luminosity and the efficiency at which the accreted mass is converted to
radiation. This requirement ensures that the disk can remain in a steady
state, capturing the mass it needs to continue fueling the black hole,
and will lead to a relation between the black hole mass and the velocity
dispersion of the stellar system.

  This still leaves two remaining questions: how the accretion disk is
initially started so that it can gain matter by capturing stars, and
when the mechanism of stellar captures and black hole growth stops. We
will propose that when a starburst takes place in a galactic nucleus, a
small seed black hole with an initial accretion disk (made, for example,
by tidally disrupted stars) naturally grows its accretion disk by
capturing stars, and maintains it at the level required to accrete and
shine as a quasar with a luminosity near the Eddington value. The black
hole grows in mass until it reaches the value determined by
our derived relation with the velocity dispersion of the stars that
were formed around the black hole. After this value is reached, the rate
of star captures is too small to maintain an accretion rate near the
Eddington value. This reduces the surface density of the disk, further
decreasing the star capture rate until, perhaps when accretion switches
from a thin disk to an advection mode, the quasar turns off and leaves
the black hole at a fixed predetermined mass. This can then explain the
observed $M_{BH}-\sigma_*$ relation of the present remnant black holes
with the velocity dispersion of the spheroidal systems around them.

  We start by summarizing the standard steady-state thin accretion
disk model, transporting angular momentum by an effective viscosity as
described in the usual $\alpha$-model (Shakura \& Sunyaev 1973;
Pringle 1981). We closely follow Goodman (2003), reproducing some
equations here for completeness. In order to determine whether a star
that plunges through the disk can be captured, we must first
obtain the disk surface density profile.

\subsection{The Surface Density Profile of Steady Accretion Disks}

  Matter in an accretion disk can move in towards the black hole as
angular momentum is transported out by viscous processes. We assume a
steady-state accretion disk of surface density $\Sigma(r)$ at radius $r$
and angular rotation rate $\Omega=(GM_{BH}/r^3)^{1/2}$, with a constant
accretion rate $\dot M$, where viscosity causes the gas to drift inward
at a radial speed $v_r(r)$ which is much smaller than the tangential
orbital velocity $\Omega r$. The viscosity force per unit area can be
imagined as acting on the surface of a cylinder at radius $r$ that cuts
the disk vertically, and is equal to $\nu \rho r \Omega'$, where $\nu$
is the viscosity coefficient, $\rho$ is the density, and $\Omega' =
d\Omega/dr$. This force, integrated vertically, causes a torque on the
disk inside the cylinder of $2\pi r \nu \Sigma r^2 \Omega'$. Considering
a ring of the disk at radius $r$, the net torque acting on the ring per
unit ring width $dr$ is $d(2\pi \nu \Sigma r^3 \Omega')/dr$. The angular
momentum of the ring per unit ring width is $2\pi r \Sigma \Omega r^2$,
and conservation of angular momentum in steady-state conditions implies
\begin{equation}
{d (\Sigma r^3 \Omega v_r) \over dr} = {d (\nu \Sigma r^3 \Omega' )
\over dr } ~. 
\end{equation}
Integrating this equation, and substituting $\Omega' = - 3 \Omega/(2r)$,
and $v_r = - \dot M / ( 2\pi r \Sigma )$
as required by mass conservation in steady-state, we have
\begin{equation}
\nu \Sigma = { \dot M \over 3\pi } \left[ 1 - \left( r_{int} \over r
\right)^{1\over 2} \right] ~,
\end{equation}
where $r_{int}$ is an integration constant that depends on an inner
boundary condition. In practice, $r_{int}$ is determined by the
relativistic inner regions of the accretion disk and is negligible in
the outer regions that we will be interested in, so we can use
\begin{equation}
\Sigma = { \dot M \over 3\pi \nu } ~.
\label{sdpf}
\end{equation}

  The viscosity is usually assumed to be related to turbulent processes,
and is of the order of the product of the velocity and size of the
largest turbulent eddies. Following Goodman (2003), the viscosity
coefficient is expressed as $\nu = \alpha \beta^b\, c_s^2/\Omega$, where
$c_s = \sqrt{p/\rho}$ is the isothermal sound speed at the midplane,
$\rho$ is the gas density, $\beta= p_{gas}/p$, and $p_{gas}$ and $p$ are
the gas and total pressure, respectively, with $p-p_{gas} = p_{rad}$
being the
radiation pressure. If viscosity is produced by magnetorotational
instability, the dimensionless viscosity parameter $\alpha$ is thought
to be between $10^{-3}$ and $10^{-1}$ (Balbus \& Hawley 1998). When
radiation pressure dominates, we assume the viscosity may be
proportional to either the total pressure ($b=0$) or to the gas pressure
($b=1$).

  Writing also $\dot M = L/(\epsilon c^2) = \dot m\, L_{Edd}/c^2$,
where $L$ is the radiative luminosity of the disk, $\epsilon$ is the
radiative efficiency, $L_{Edd} = 4\pi c G M_{BH}/\kappa_e$ is the
Eddington luminosity, $\kappa_e$ is the electron scattering opacity, and
the accretion rate has been conveniently parameterized as $\dot m =
L/(\epsilon L_{Edd})$, we obtain
\begin{equation}
\dot M = { 4 \pi G M_{BH} \dot m \over c \kappa_e\, } ~,
\label{dotm}
\end{equation}
and
\begin{equation}
\Sigma = { 4 G M_{BH} \dot m \Omega \over 3 \alpha c \kappa_e\,
c_s^2 \beta^b } ~.
\label{sdps}
\end{equation}

  To proceed further, it is necessary to specify the energy balance in
the disk to compute the midplane temperature, $T$, related to the
isothermal sound speed by $\beta c_s^2 = (k_B T) / (\mu m_p)$, where
$k_B$ is the Boltzmann constant and $\mu m_p$ is the mean particle mass.
Assuming that viscous dissipation of the orbital energy is the dominant
heat source, the energy dissipation rate per unit area of the disk is
$\nu \Sigma (r\, \Omega')^2$. This must be equal to the radiative energy
emitted per unit area by the two sides of the disk, $2\sigma T_{eff}^4$,
where $\sigma$ is the Stefan-Boltzmann constant. Substituting $\Omega' =
-(3/2) (GM_{BH})^{1/2} / r^{5/2}$ and using equations (\ref{sdpf}) and
(\ref{dotm}), one finds
\begin{equation}
\sigma T_{eff}^4 = { 3G^2M_{BH}^2\, \dot m \over 2 c \kappa_e r^3} ~. 
\end{equation}
The midplane temperature can be approximated by $T^4\simeq \kappa\,
(\Sigma/2)\, T_{eff}^4$, where $\kappa$ is the opacity. Hence,
\begin{equation}
T = \left( { 3G^2M_{BH}^2\, \dot m\, \hat\kappa \Sigma \over
4\sigma c\, r^3} \right)^{1\over 4} ~, 
\label{midtemp}
\end{equation}
where $\hat\kappa = \kappa/\kappa_e$. Replacing the temperature by
the isothermal sound speed, substituting into equation (\ref{sdps}), and
using $\sigma = (2\pi^5 k_B^4)/(15 c^2 h^3)$, where $h$ is the Planck
constant, we find
\begin{equation}
\Sigma = { 4 \pi \over 3c} \left[2 \over 15 \hat\kappa(r)
\right]^{1\over 5} \, \left(\dot m \over h \, r \right)^{3\over 5}
\, \left[ G M_{BH} \mu m_p \over \alpha \kappa_e\, \beta(r)^{b-1}
\right]^{4\over 5} ~.
\label{sdpt}
\end{equation}
The quantity $\beta=p_{gas}/p$ can be expressed in terms of $\Sigma$ and
$T$ using $p_{gas}=\rho k_B T/(\mu m_p)$, $p_{rad}=4\sigma T^4/(3c)$,
and $\rho=\Sigma\Omega/(2c_s)$ (since the disk scale height is
$H = c_s/\Omega$). Using equations (\ref{midtemp}) and (\ref{sdpt}),
one finds
\begin{equation}
{\beta^{4+b \over 10}\over 1-\beta} = { (45)^{1\over 10}\,
\pi^{3\over 40} \over 2^{1\over 20} }\,\left(
m_{Pl}^2 m_e^2 \over \alpha \alpha_e^2 \mu^4 m_p^3 M_{BH} \hat\kappa^9
\right)^{1\over 10}\, \dot m^{-4\over 5} \,
\left(r\over R_S\right)^{21\over 20} ~.
\end{equation}
Here, we have used $\kappa_e = 4\pi \alpha_e^2 \hbar^2 (1+X) / (3m_p
m_e^2 c^2)$, where $\hbar=h/(2\pi)$, $m_p$ is the proton mass, $X$ is
the hydrogen abundance by mass, and $\alpha_e$ is the fine structure
constant. We also use the Schwarzschild radius $R_S=2GM_{BH}/c^2$, and
the Planck mass $m_{Pl}=\sqrt{\hbar c/G}$. For characteristic values of
the black hole mass $M_{BH}= 10^{8} M_8 \msun$, $\alpha = 10^{-2}
\alpha_{-2}$, $X=0.7$, and $\mu = 0.62$, the $\beta$ parameter is given
by
\begin{equation}
{\beta^{4+b \over 10}\over 1-\beta} = 4.6 \alpha_{-2}^{-{1\over 10}}\,
\hat\kappa^{-{9\over 10}}\, \dot m^{-{4\over 5}} \, M_8^{-{1\over 10}}\,
\left(r\over 10^3 R_S\right)^{21\over 20} ~,
\label{betas}
\end{equation}
the surface density is
\begin{equation}
\Sigma = 7.6\times 10^5\, \left( \alpha_{-2} \beta^{b-1}
\right)^{-{4\over 5}} \, \dot m^{3\over 5}\,
\left( {M_8 \over \hat\kappa} \right)^{1\over 5} \,
\left( {10^3 R_S \over r} \right)^{3\over 5} \gm\cm^{-2} ~,
\label{sdpq}
\end{equation}
and the midplane temperature is
\begin{equation}
T = 3.9\times 10^4\, {\rm K} \left( \hat\kappa \over \alpha_{-2}
\beta^{b-1} M_8 \right)^{1\over 5} \, \dot m^{2\over 5}\, 
\left( {10^3 R_S \over r} \right)^{9\over 10} \kel ~.
\end{equation}

  We plot in Figure 1 the contours in the $M - R/R_S$ plane at which
$\beta=0.5$, and at which the surface density and midplane temperature
have the characteristic values indicated in the caption, for the case
$\dot m = 10$ and $\alpha_{-2}=1$. For all black hole masses, the
pressure in the accretion disk is dominated by
radiation to the left of the line $\beta=0.5$, and is dominated by
matter to the right of this line. For the case $b=1$, $\Sigma$ is
independent of $\beta$ and, assuming that $\hat\kappa(r)\simeq 1$ and
$\alpha$ is independent of $r$, then $\Sigma(r)\propto M_{BH}^{4/5}/
r^{3/5}$. This form of the surface density profile will be important
later for our $M_{BH}-\sigma_*$ relation. We also show the contour for
which the opacity due to bound-free transitions is $\kappa_{bf} =
\kappa_e$ (implying $\hat\kappa=2$ if other opacity mechanisms are
negligible). To the left of this line we have $\hat\kappa\simeq 1$.

\subsection{Changes of Velocity, Orbital Energy and Mass of a Plunging
Star}

  The orbital angular momentum of a star in the stellar system around
the black hole may be changed by relaxation processes or a triaxial
potential (due to the gravitational contribution of stars and the gas
disk), and if by chance the orbit becomes very eccentric the star may
collide with the dense central part of the accretion disk. The gas drag
force during the collision will then slow down the star by a velocity
increment $\Delta v$, resulting in a loss of orbital energy and a
corresponding reduction of the
orbital apocenter and period. If the apocenter reduction is substantial
enough, the star will be condemned to plunge repeatedly through the
disk until its orbit is brought into the disk plane and circularized
(Syer \etal 1991), leaving the star immersed inside the disk. This
defines an effective loss-cylinder for the capture of a star by the
disk. We now calculate the condition for the star to be captured.

  As we shall see later, the typical radius at which stars are captured
by crossing the disk is $r\sim 10^3 R_S$, where $R_S$ is the
Schwarzschild radius of the black hole. The orbital velocity of the
star as it crosses the disk at this radius is $\sim 10^4 \kms$, which is
highly supersonic and much greater than the escape velocity from the
surface of the star. This means that gravitational focusing is
negligible, and hence the response of the disk to the passage of the
star can be ignored for the purpose of computing the velocity change of
the star $\Delta v$.
Under these conditions we assume, as a first approximation, that
$\Delta v$ is simply determined by the absorption of the momentum of all
the disk gas that lies along the path of the star:
\begin{equation}
\Delta v = {\Sigma \over \Sigma_*}\, { v \over \sin\theta } ~,
\label{Dv}
\end{equation}
where $\Sigma_* = M_*/(\pi R_*^2)$ is the mean surface density of the
star of mass $M_*$ and radius $R_*$, $\theta$ is the angle between the
orbital velocity of the star and the plane of the disk at the
intersection point, and $v$ is the relative velocity between the star
and the disk gas moving on a circular orbit. Typically, $\Sigma_* \sim
10^{11} \gm\cm^{-2}$ and $\Sigma \sim 10^6 \gm\cm^{-2}$ (see Fig.\ 1),
so $\Delta v/ v\sim 10^{-5}$ for each star passage through the disk.

  Apart from reducing their velocity, stars may also lose a small
fraction of their mass at every disk passage. As will be discussed below
in \S 4, stars may survive essentially intact through many disk
passages until they merge inside the disk, but after this they should
be destroyed and their matter should dissolve into the disk.


  A complete analysis of the rate at which stars are captured by the
disk would require computing the probability distribution of $\Delta v$
from a random distribution of orbital inclinations and pericenter
longitudes (see Ostriker 1983 for a calculation along this line to
compute the rate of angular momentum loss of the disk by star
crossings). In this paper we make a more approximate estimate based on
considering a typical star orbit. We use $v\simeq (2GM_{BH}/r)^{1/2}$
(neglecting the circular velocity of the gas in the disk), and assume
the case where the star crosses the disk at its pericenter. The average
value of $1/\sin\theta$, where $\theta$ is the angle between the plane
of the orbit and the plane of the disk (with probability distribution
$\sin\theta\, d\theta$), is then $\pi/2$, so the change in the orbital
energy per unit mass is
\begin{equation}
\Delta E = v \Delta v = {\Sigma\over \Sigma_*}
{ \pi G M_{BH}\over r_p} ~,
\label{eloss}
\end{equation}
where $r_p$ is the pericenter. Averaging over all possible pericenter
longitudes would not greatly modify this result (for example, one can
easily show that for a pericenter longitude of 90 degrees, the collision
of the star with the disk takes place at a radius $2r_p$, which reduces
the disk surface density by $2^{-3/5}$ and the square of the relative
velocity by $1/2$, but then $\Delta v$ is increased by a factor 2
because there are two equal collisions per orbit and by another factor
$\sqrt{2}$ because the sine of the angle between the disk plane and the
stellar velocity is smaller by a factor $\sqrt{2}$).

\subsection{The Size of the Region from which Stars are Captured}

  We now consider that we have a stellar system that is roughly
isotropic in the central region of the galaxy where the gravitational
force is dominated by the nuclear black hole. Among all orbits with a
semimajor axis $a$, the probability distribution of the orbital
pericenter $r_p$ is $2(1-r_p/a)\, dr_p/a$ (because an isotropic system
in a point-mass potential has a flat distribution in the square of the
eccentricity). Hence, if the time during which the gas disk is
maintained with a roughly constant surface density is $t_S$ (which we
identify later with the Salpeter time for the growth of a black hole,
see Salpeter 1964), the number of times that a star will cross the disk
at a pericenter smaller than $r_p$ is $N_{orb} = (t_S/P)
\, (r_p/a)\, (2-r_p/a)$, where $P=2\pi a^{3/2} / (GM_{BH})^{1/2}$ is the
period. Using equation (\ref{eloss}), the average rate of orbital energy
loss of a star with semimajor axis $a$ due to passages at all
pericenters $r_p$ is
\begin{equation}
{dE\over dt} = \int_{r_1}^a {dr_p \over aP}\, 2 \left( 1-{r_p\over a}
\right) {\Sigma(r_p) \over \Sigma_*}\, {\pi G M_{BH}\over r_p} ~.
\end{equation}
Note that this integral diverges at small $r_p$, that is to say, the
loss of orbital energy is dominated by the disk crossings closest to the
center, when the eccentricity is closest to one. We choose the lower
limit of the integral, $r_1$, to be the pericenter for which
$N_{orb}=1$, because a typical star will not cross the disk at a smaller
pericenter. Using the approximation $r_1 \ll a$, this pericenter is $r_1
= (aP)/(2t_S)$. Note also that it is essential here to assume that the
rate of relaxation is fast enough to keep the loss-cylinder full, so
that the probability to have a star in a pericenter range $dr_p$ (when
$r_p \ll a$) is $2 dr_p/a$. Assuming $\Sigma(r) \propto r^{-3/5}$ (which
is true for $b=1$, or for the gas-pressure-dominated region of the disk
if $b=0$), we have
\begin{equation}
{dE\over dt} = {10 \over 3P}\,
{\Sigma(r_1) \over \Sigma_*}\, {\pi G M_{BH}\over a} ~.
\end{equation}
Substituting the orbital energy per unit mass in a point-mass potential,
$E=GM_{BH}/(2a)$, we find
\begin{equation}
{da\over dt} = - {20 \pi a \over 3P}\, {\Sigma(r_1) \over \Sigma_*} ~.
\label{dadt}
\end{equation}

  Assuming again $\Sigma(r_1) \propto r_1^{-3/5}$, and using $P\propto
a^{3/2}$, and $r_1 \propto aP \propto a^{5/2}$, we have $da/dt \propto
a^{-2}$. Hence, the condition for the star to be captured by the
disk in a time less than $t_S$ is that, at the initial semimajor axis,
\begin{equation}
\| da/dt \| > a/(3t_S) ~.
\label{dadtc}
\end{equation}

  Substituting (\ref{dadtc}) into equation (\ref{dadt}),
we find that the capture
condition is that the initial semimajor axis must be smaller than a
critical radius $r_c$ that obeys
\begin{equation}
{\Sigma(r_{1c})\over \Sigma_*} = {P_c \over 20\pi t_S} = {r_c^{3/2}
\over 10 (GM_{BH})^{1/2}\, t_S } ~,
\label{capcon}
\end{equation}
where $P_c$ is the period at $r_c$, and $r_{1c} = r_c P_c/ (2t_S)$.

  Because the loss of orbital energy is dominated by disk crossings at
the smallest pericenters, a typical star starting at $a\simeq r_c$
loses little orbital energy until by chance it enters the loss-cylinder
at a pericenter $\lesssim r_{1c}$, and then substantially reduces its
semimajor axis in a single disk crossing. After that, relaxation may
take the star out of the loss-cylinder if it occurs fast enough,
but as $a$
decreases the timescale for randomly entering the loss-cylinder again is
rapidly shortened in any case (because on average, $da/dt \propto
a^{-2}$ as shown above), so the star will inevitably be captured. The
radius $r_c$ represents the semimajor axis for which the average time to
enter the loss-cylinder for disk capture is $t_S$.

  In practice, the rate of relaxation will not be so large to make all
stars of a fixed semimajor axis have the same probability to be
captured. Stars on loop orbits may never come close to the center, and
only the stars that can reach the phase space region of zero angular
momentum from their initial orbit with the available rate of relaxation
will be captured. Provided these latter stars are a large enough
fraction of the total within the radius $r_c$, the approximation of a
full loss-cylinder is appropriate and equation (\ref{dadt}) still gives
an average rate of reduction of the semimajor axis. Some stars on loop
orbits within $r_c$ may never reach the loss-cylinder, while other stars
slightly further than $r_c$ may be captured if they start on highly
radial orbits. We will ignore these complications here and assume that
the stars that are captured are those that start with $a< r_c$.

\subsection{The Black Hole Mass - Velocity Dispersion Relation}

  We can now clearly see why there should be a relation between the
black hole mass and velocity dispersion of the stellar system around the
black hole. Assuming the stars follow an isothermal density profile,
the stellar system
has a mass within a radius $r_c$ given by its velocity dispersion
$\sigma_c$, $M_*(r_c) \simeq 2 r_c \sigma_c^2/G$. Over the time duration
$t_S$ of the luminous phase of a quasar during which the gaseous disk
is present, stars will continuously plunge through the disk supplying it
with new gas, which will then accrete into the black hole.
At the end of this process, the black hole mass will be $M_{BH} =
M_*(r_c)$, so $r_c$ is equal to the zone of influence of the black hole
as its mass reaches its final value.

  From equation (\ref{capcon}), the surface density at the pericenter
where stars are captured, $\Sigma(r_{1c})$, is proportional to the
orbital period $P_c$, if $t_S$ and $\Sigma_*$ are constant. Since
\begin{equation}
 r_{1c} \propto r_c P_c \propto r_c^{5/2}/M_{BH}^{1/2} ~,
\end{equation}
and, for $b=1$ and $\hat\kappa = 1$, we have $\Sigma(r) \propto
M_{BH}^{4/5}/r^{3/5}$ [eq.\ (\ref{sdpt})], we find
\begin{equation}
\Sigma(r_{1c}) \propto M_{BH}^{4/5}/r_{1c}^{3/5} \propto
M_{BH}^{11/10}/r_c^{3/2} \propto P_c \propto
r_c^{3/2}/M_{BH}^{1/2} ~.
\end{equation}
This implies that $M_{BH}^{8/5} \propto r_c^3$, or, using
$r_c\propto M_{BH}/\sigma_c^2$, that
\begin{equation}
M_{BH} \propto \sigma_c^{30/7} ~, 
\end{equation}
in excellent agreement with the slope of the observed power-law relation
(Merritt \& Ferrarese 2001, Tremaine \etal 2002).

  The full predicted relation between $M_{BH}$ and $\sigma_c$ is easily
found starting from equations (\ref{sdpt}) and (\ref{capcon}), combined
with $r_{1c}=\pi r_c^{5/2}/[t_S (GM_{BH})^{1/2}]$. We also use the
relation $r_c = GM_{BH}/(2\sigma_c^2)$ for an isothermal density
profile, and $t_S = (c \kappa_e)/(4\pi G \dot m)$ for the Salpeter
time. The result is
\begin{equation}
M_{BH} = M_*^{5\over 7}\, m_{Pl}^{2\over 7}\,
\left(\alpha_e m_{Pl}\over m_e \right)^{8\over 7} 
\left(2\sigma_c\over v_{e*} \right)^{30\over 7} 
\left( v_{e*}\over 3c \right)^{10\over 7} \,
\dot m^{-{5\over 7}} \, \hat\kappa(r_{1c})^{-{1\over 7}}
\left[ 5\mu (1+X) \over \alpha \beta(r_{1c})^{b-1} \right]^{4\over 7} ~,
\label{msir}
\end{equation}
where $m_{Pl} = (\hbar c/G)^{1/2}$ is the Planck mass and $v_{e*} =
(2GM_*/R_*)^{1/2}$ is the escape velocity of the star.
In terms of typical values of $M_{BH}$ and $\sigma_c$, this is
\begin{equation}
M_{BH} = 3.3\times 10^8 \msun \, \left( M_*\over \msun \right)^{5\over 7}
\left(\sigma_c \over 200 \kms \right)^{30\over 7}
\left(v_{e\odot} \over v_{e*} \right)^{20\over 7} \,
\dot m^{-{5\over 7}} \, \hat\kappa(r_{1c})^{-{1\over 7}}
\left[ \alpha_{-2} \beta(r_{1c})^{b-1} \right]^{-{4\over 7}} ~,
\label{msis}
\end{equation}
where we have used $X=0.7$ and $\mu = 0.62$.

  The normalization of the relation is also reasonably close to the
observed relation $M_{BH} = 1.3\times 10^8 \msun (\sigma_*/200 \kms)^a$
(Merritt \& Ferrarese 2001, Tremaine \etal 2002), but it is subject to
many uncertainties depending on various parameters:
the type of star that is most abundant in the stellar system, the
viscosity parameter $\alpha$, the normalized accretion rate $\dot m$,
and the relationship of the velocity dispersion $\sigma_c$ of the
nuclear starburst region at the time the quasar was active to the
present central velocity dispersion of the galaxy that contains the
remnant black hole. On this last point, it must be realised that
the stellar density we have assumed at radius $r_c$ is much higher than
the present stellar density in observed ellipticals (for example, for
$\sigma_c=200 \kms$ and $M_{BH}\sim 10^8 \msun$, $r_c\simeq 5$ pc, and
the density profiles of observed ellipticals are shallower than
isothermal up to radii much larger than $r_c$). This implies that in
our model the central stellar density must have been greatly reduced
after the quasar epoch due to, for example, mass loss from supernovae
and winds with subsequent adiabatic expansion (in particular if the
stellar mass function in nuclear starbursts is dominated by massive
stars), and mergers of nuclear black holes as their host elliptical
galaxies merge (Quinlan 1996; Faber \etal 1997; Milosavljevi\'c \&
Merritt 2001). These processes may
have altered the central velocity dispersion of the galaxy.

  The quantities $\hat\kappa(r_{1c})$ and $\beta(r_{1c})^{b-1}$ can
affect not only the normalization but also the shape of the
$M_{BH}-\sigma_c$ relation. In Figure 1, the value of $r_{1c}$ as a
function of $M_{BH}$ is plotted (assuming $\hat\kappa=1$). This is
computed by using $r_{1c} = r_c P_c/(2t_S)$, which yields:
\begin{equation}
r_{1c} = 340 R_S\, \dot m^{1\over 6}\,
\left(M_*\over M_{\odot} \right)^{5\over 6}\,
\left( v_{e\odot}\over v_{e*} \right)^{10\over 3}\,
(M_8 \hat\kappa )^{-{1\over 6}}\, (\alpha_{-2}\beta^{b-1})^{2\over 3} ~.
\label{r1cf}
\end{equation}
As we can see, the region where $\hat\kappa$ exceeds unity substantially
is at radius larger than $r_{1c}$, but the disk radiation pressure can
be important near $r_{1c}$. This implies that the relation $M_{BH}
\propto \sigma_*^{30/7}$ is only valid for $b=1$, that is to say, when
the disk viscosity is proportional to the gas pressure. If $b=0$, and in
the radiation-dominated pressure limit of $\beta \ll 1$, we can find by
using equations (\ref{betas}), (\ref{msis}), and (\ref{r1cf}) that
$M_{BH} \propto \sigma_c^{10/3}$. Hence, the predicted $M_{BH}-\sigma_c$
relation depends on details of the disk viscosity.

  Generally, the $M_{BH}-\sigma_c$ relation in (\ref{msis}) depends on
the disk surface density profile, which is subject to other possible
modifications in addition to the viscosity mechanism. The effects of the
stars crossing the disk on the surface density profile will be briefly
discussed in \S 5. The simple thin disk model may also modified when
the luminosity is approximately equal to the Eddington luminosity, and
the disk thickens by the radiation pressure
(Wang \etal 1999; Ohsuga \etal 2002).

\subsection{The Evolution of a Nuclear Black Hole after the Formation of
a Galactic Spheroid}

  The mass-velocity dispersion relation we have derived in equation
(\ref{msir}) originates from the condition that the rate at which new
mass is delivered to the disk by plunging stars is the same as the
rate at which the gas in the disk is accreted by the black hole.
How does that establish the black hole mass-velocity dispersion relation
in the remnant galaxies after quasars turn off?

  Mergers of galaxies rich in cold gas may often give rise to strong
nuclear starbursts. A seed black hole may be present in the nucleus,
probably coming from one of the galaxies that merged. The seed black
hole mass can initially be very small compared to the value implied by
equation (\ref{msir}) for the velocity dispersion of the newly formed
galactic spheroid. A small seed accretion disk can initially form
around this black hole by a variety of mechanisms, such as residual gas
left over from star formation or from stellar winds that reaches the
center directly, physical stellar collisions, and tidal disruptions of
stars by the black hole. This disk can then start growing and feeding
the black hole by capturing stars. We now show that the capture rate of
stars will be more than sufficient to maintain an Eddington accretion
rate as long as the black hole mass is below the $M_{BH}-\sigma_c$
relation in (\ref{msir}).

  For fixed $\sigma_c$ and a singular isothermal profile of stars, the
black hole needs to capture stars out to a radius $r_c \propto M_{BH}$
to increase its own mass. Let t be the timescale over which these stars
with total mass $M_{BH}$ are captured. Then, as the black hole grows the
capture pericenter must increase as $r_{1c} \propto r_c P_c/t \propto
M_{BH}^2/t$ (where the orbital period at $r_c$ is $P_c \propto
r_c^{3/2}/M_{BH}^{1/2} \propto M_{BH}$). By the same arguments as in the
previous subsections, capturing a star over one orbit requires
$\Sigma(r_{1c})/\Sigma_* \propto r_{1c}/r_c \propto M_{BH}/t$.
At the same time, if accretion of the disk gas to the black hole also
occurs at the rate $\dot M \sim M_{BH}/t$ (so the normalized accretion
rate is $\dot m \propto t^{-1}$),
then equation (\ref{sdpt}) implies that the surface density at
the capture radius is $\Sigma(r_{1c})\propto M_{BH}^{4/5} \dot m^{3/5} /
r_{1c}^{3/5} \propto M_{BH}^{-2/5}$, independent of $t$. Combining these
two scalings for $\Sigma(r_{1c})$, we infer $t\propto M_{BH}^{7/5}$.
Equivalently, the normalized accretion rate is $\dot m\propto
M_{BH}^{-7/5}$. This makes it clear that the lower the mass of the black
hole, the shorter the time required for the gas disk to capture stars
with a mass equal to that of the black hole. And viceversa, if the black
hole exceeds the mass in equation (\ref{msir}), then the gas disk must
take longer than the time $t_S$ to capture stars with a total mass
$M_{BH}$.

  We note that our inferred proportionality $\dot m \propto
M_{BH}^{-7/5}$ when the disk is capturing stars from a stellar cluster
with fixed $\sigma_c$ implies that the accretion rate is super-Eddington
when the black hole mass is small. Under these circumstances, a thin
disk with the surface density profile (\ref{sdpt}) is probably not valid
(although disk models with super-Eddington accretion have been proposed
by Begelman 2001). Nevertheless, this does not alter our conclusion that
stellar capture provides sufficient matter for maintaining an accretion
rate at least at the Eddington level while the black hole mass is below
the relation (\ref{msir}). For example, if we assume that a thin disk
with fixed $\dot m$ is maintained while the black hole mass is small
(with the ``excess'' matter delivered by stars being expelled in an
outflow), then we have $\Sigma(r_{1c})\propto M_{BH}^{4/5}/r_{1c}^{3/5}
\propto M_{BH}^{-2/5} t^{3/5}$, which combined with $\Sigma(r_{1c})
\propto M_{BH}/t$ yields $t\propto M_{BH}^{7/8}$, so the timescale for
delivery of a mass $M_{BH}$ to the disk is still an increasing function
of $M_{BH}$.

  Hence, capture of stars will replenish the gas disk at a rate fast
enough to maintain an Eddington accretion rate (with $L\simeq L_{Edd}$
and $\epsilon \simeq 0.1$) while the black hole mass is less than in
equation (\ref{msir}) for $\dot m \simeq 10$. When this value of the
black hole mass is reached, the disk may still continue to be fed at a
decreasing value of $\dot m$, and the black hole mass can keep growing
up to a value $M_{BH} \propto (\dot m)^{-5/7}$, according to equation
(\ref{msir}). This continued growth requires that star captures continue
over the increasingly long timescale $t_S \propto \dot m^{-1}$.

  The growth of the black hole may eventually stop due to three possible
mechanisms. The first is the aging of the stellar population, which
reduces the number of available stars for capture and the mean value of
$M_*^{5/7}/v_{e*}^{20/7}$ in equation (\ref{msir}). The second is the
depletion of the loss-cylinder as the post-starburst stellar system
reaches equilibrium and the rate of orbital relaxation declines. The
third is a transition of the gas accretion into the black hole from a
thin, radiatively efficient disk into a hot, advection dominated
accretion flow or inflow/outflow solution (e.g., Blandford \& Begelman
1999; Quataert 2003). This transition to a hot, thick flow may occur at
a critical value of $\dot m$ as it decreases (Begelman \& Celotti 2004),
and it implies a dramatic decline of the surface density which would
immediately terminate the capture of stars and cause a sudden decrease
in $\dot m$. Therefore, our model also suggests a mechanism by which
quasars turn off, leaving behind a remnant black hole that will stay at
a constant mass thereafter if there are no additional nuclear
starbursts. This final black hole mass should therefore depend on the
value of $\dot m$ at which the accretion ceases. If this value of
$\dot m$ depended on $M_{BH}$, the slope of the $M_{BH}-\sigma_c$
relation would be altered.

  There are a number of possible problems with the simple picture
presented here that will need to be addressed in future work. The
initial process by which the gas disk grows around a seed black hole,
reaching the Eddington accretion rate out to a certain radius, needs to
be examined more carefully. The structure of the accretion disk should
also be thoroughly modified from that of the simple steady-state model
in \S 2.1, due to the effects of the capture of stars in adding mass,
energy and angular momentum to the disk. Finally, the process by which
a quasar would turn off as the stellar capture rate decreases and the
stellar population ages also needs to be studied.

\section{Self-Gravity of the Accretion Disk}

  One of the classic problems encountered by standard quasar accretion
disk models is that if matter
is transported from large radius, from any residual gas in the galaxy
that is left over from star formation, the gas should form a thin disk
and become gravitationally unstable to form stars while it is still very
far from the black hole (Shlosman \& Begelman 1987). In this case, the
result might not be a black hole fed by an accretion disk, but simply a
dense inner disk of stars. The criterion for gravitational
instability is obtained from Toomre's parameter, $Q= (c_s \Omega)/
(\pi G \Sigma)$, which using equations (\ref{midtemp}) and (\ref{sdpt})
can be expressed as (see also the Appendix in Goodman 2003)
\begin{equation}
Q = {45^{3\over 10} \over \pi^{9\over 10} 2^{27\over 20} }
\dot m^{-{2\over 5}} \hat\kappa^{3\over 10}
\beta^{7b-12\over 10} \left( m_{Pl}\over \mu m_p \right)^{6\over 5}
\left( {m_{Pl}\over M_{BH}} \right)^{13\over 10}
\left[ \alpha_e^2 \alpha m_{Pl}^3 (1+X) \over m_p m_e^2
\right]^{7\over 10} \left( R_S \over r \right)^{27\over 20} ~, 
\end{equation}
or in terms of fiducial values,
\begin{equation}
Q = 0.33 \, \dot m^{-{2\over 5}}\,
\hat\kappa^{3\over 10}\, \beta^{7b-12\over 10}\,
M_8^{-{13\over 10}}\, \alpha_{-2}^{7\over 10} \, 
\left( 10^3 R_S \over r \right)^{27\over 20} ~. 
\end{equation}
Contours of $Q=1$ are shown in Figure 1 (for $\dot m =10$,
$\alpha_{-2}=1$). To the left of the contour,
$Q>1$ and the disk is stable, while to the right the disk is
gravitationally unstable. Figure 1 shows that the region of the disk
within $r_{1c}$ is mostly stable, except for large black hole masses
for which the disk may be marginally unstable close to $r_{1c}$. The
effect of the stars crossing the disk would probably be important here
in increasing the disk temperature and stabilizing it. Hence, the model
proposed here may solve the problem of instability in quasar accretion
disks at large radius, simply because most of the matter is delivered
to the disk at small radius. Future work to calculate the disk structure
including the effects of the plunging stars and possible extensions of
the accretion disk to large radius with matter from stellar winds will
be needed to examine this question in more detail.

\section{The Fate of the Captured Stars}

  The orbits of the stars plunging through the disk are gradually
circularized and brought into the disk plane in successive crossings. At
every crossing, a fraction of the star might be stripped from its
outer layers. If this mass loss were large enough stars could be
destroyed before they merge into the disk.

  Considering a typical case, for $M_{BH}=10^8 \msun$, $\dot m =10$,
$\alpha=10^{-2}$, and solar-type stars, a star crossing the disk at
$r_{1c}\simeq 500 R_S$ [eq.\ (\ref{r1cf})] encounters a surface density
$\Sigma\sim 5\times 10^6\gm\cm^{-2}$ with midplane
temperature $T\sim 2\times 10^5$ K, $\beta\sim 0.4$, and $c_s\sim 70
\kms$ (as obtained from \S 2.1). The disk scaleheight is $H=c_s\Omega
\sim 6$ AU, and at a velocity $v\sim 10^4 \kms$ the star takes a few
days to cross the disk. The midplane disk density at $r_{1c}$ is
$\rho\sim
\Sigma/(2H)\sim 10^{-7.5} \gm\cm^{-3}$. The ram pressure exerted by
the disk gas on the star is $\rho v^2 \sim 10^{10.5} \erg\cm^{-3}$. In
a star like the Sun, this same pressure occurs at a radius $0.97 R_*$,
which corresponds to a mass $M_{out} \sim 10^{-4.5} M_*$ outside this
radius. However, most of this mass should simply be pushed against the
star surface to higher pressure by the shock generated by the disk
wind, and will not be lost. The disk material will be heated to $\sim
10^9$ K when it encounters the shock, and will then flow around the
star. Any stellar mass loss during the disk passage depends on the
heat conduction and irradiation from the post-shock disk material into
the stellar surface, and any turbulent instabilities that may mix
stellar matter into the disk wind. After the disk passage, the heated
material in the stellar surface would cool back to equilibrium on a
short timescale compared to the orbital period.  We therefore expect
that the mass loss of the plunging stars will be negligibly small,
modulo the uncertain mixing processes of matter in the stellar surface
and the post-shock disk wind that might act to destroy stars before they
merge into the disk (see also Goodman \& Tan 2003).

  Assuming that the stars survive, once embedded inside the disk they
may create gaps around them if they are sufficiently massive, or they
may accrete gas from the disk (Syer \etal 1991, Artymowicz \etal 1993).
Stars might also evaporate into the disk instead of accreting if the
midplane temperature is high enough. In any case, eventually the stars
will be mostly destroyed and dissolved into the gaseous disk. That
this must be the case can easily be seen by considering what would
happen if a large fraction of the disk mass were in the form of stars of
$M_*\simeq 1\msun$. For typical parameters $M_8=1$, $r=500 R_S = 1000
{\rm AU}$, and $\Sigma = 10^{6.5} \gm \cm^{-2}$, stars would have to
occupy a fraction of the disk area of at least $10^{-5}$ (given their
typical surface density $\Sigma_*\sim 10^{11}\gm\cm^{-2}$), so in a ring
of width equal to the solar diameter there would be about 10 stars.
These numbers become even greater for more massive stars because their
surface density is lower. This makes it clear that collisions and
scatterings among stars would be frequent, and they would soon lead to
coalescence and destruction of the stars. If a star is on a slightly
eccentric orbit (either because the orbit has not yet fully circularized
after capture or because it has been scattered by other stars), a
collision can take place at a sufficiently high relative velocity to
cause the dissolution of the stars rather than coalescence. If stars can
coalesce repeatedly, they will become very massive and lose a lot of
their mass in winds (see Goodman \& Tan 2003 for a discussion of the
possible presence of supermassive stars in accretion disks). Very
massive stars are also highly unstable because their internal energy is
almost balanced by their negative gravitational energy (since they are
polytropes with adiabatic index very close to $4/3$), so collisions with
other stars would not have to occur at very high velocities to produce
large amounts of mass loss. It therefore appears inevitable that a large
fraction of the mass in the stars will be dissolved into the disk, and
can eventually be accreted onto the black hole with high radiative
efficiency.



\section{The Angular Momentum Problem}

  The idea presented in this paper is that most of the matter in the
black hole accretion disk comes from stars that were originally in
orbits with radius much larger than the size of the disk. These stars
were perturbed into highly eccentric orbits and captured by the disk,
and were embedded into the disk after repeated disk crossings in which
their orbital energy was dissipated.

  If we imagine that the accretion disk has to be initially formed from
the matter of these plunging stars (starting from an initial small disk
which can grow in mass from captured stars until it reaches a
steady-state structure), we are faced with the problem of how the disk
has acquired its angular momentum. The average angular momentum of the
captured stars is far too small to make a disk. Although the specific
angular momentum of an individual plunging star is of the same order as
the specific angular momentum of the disk gas at the radius $r_{1c}$
where the star is captured, the direction of the angular momentum of
each star is basically random. Even if the stellar system around the
black hole is rotating (implying a phase space density of stars that
depends on the direction of the orbital angular momentum), the stars
captured by the disk are coming from a very narrow loss-cylinder, and
the variation of the phase space density over this narrow region is
negligible. Thus, if the disk is made solely by these
stars, the final specific angular momentum of the captured matter is
reduced by the square root of the number of stars that have contributed
to the disk mass. In other words, the radius of the disk that could be
made by the matter from these stars is reduced relative to the capture
radius $r_{1c}$ by a factor equal to the number of captured stars, which
would make the disk smaller than the Schwarzschild radius. Clearly, a
disk around the black hole cannot be made by stars captured from
orbits with random orientations.

  This problem is, however, much less severe if the disk is considered
to be in place initially, and the plunging stars only have to maintain
the disk in a steady state. If we imagine that the disk is truncated at
a radius $\sim r_{1c}$, where most of the plunging stars
are incorporated to the disk, then the angular momentum that flows out
of the disk and into the black hole is very small (because the radius of
the innermost stable circular orbit from which matter is accreted to the
black hole is much smaller than $r_{1c}$), and the angular momentum that
flows into the disk from the plunging stars is also very small. Hence,
the disk angular momentum can basically be preserved: as matter flows
inwards in the disk, it releases its angular momentum by viscosity
processes towards the outer disk, and this angular momentum is
constantly being absorbed by the matter added from plunging stars. With
this simplified description, it can be argued that the scaling of the
disk surface density profile with black hole mass and radius in equation
(\ref{sdpt}) would essentially be preserved when the energy and angular
momentum added to the disk by the plunging stars are
included self-consistently, except that the disk is truncated around
radius $r_{1c}$ where most of the matter from stars is added.

  This greatly alleviates, but does not completely solve, the angular
momentum problem, because there is still a small amount of angular
momentum that is transferred to the black hole as the accretion
proceeds. Moreover, the problem may be made worse because the stars that
have angular momentum opposite to that of the disk should have a higher
probability of capture owing to their larger relative velocity with
respect to the disk material. This implies that the captured stars would
actually carry a net average angular momentum to the disk but with
opposite direction. A full solution of the problem may require the
addition of some matter to the disk with high specific angular momentum
from large radius. The disk would not need to be completely cut off at
radius larger than $r_{1c}$, but it could continue at a lower surface
density. The outer-disk matter with high angular
momentum can originate from gas that is left over from star formation
or has been expelled in winds from evolved stars and supernova
explosions, and its angular momentum can come from a small rotation rate
of the stellar system around the black hole.
Note that the matter accreting from a disk at radius larger than
$r_{1c}$ needs to be only a small fraction of the total, with most of
the mass coming from plunging stars (because only a small fraction
of the disk angular momentum needs to be replenished every time its mass
is replaced within $r_{1c}$), thereby preserving the explanation for the
$M-\sigma$ relation we have proposed in this paper. As for the problem
that stars moving on orbits with angular momentum opposite to that of
the disk are more likely to be captured than stars moving in the same
sense of rotation as the disk, we note that while stars
lose most of their orbital energy during disk crossings at the smallest
pericenter, the exchange of angular momentum with the disk is dominated
by crossings at large radius for a disk surface density profile
$\Sigma(r) \propto r^{-3/5}$. This may help solve the problem in a
disk extending to $r \gg r_{1c}$ with a small fraction of the accretion
rate contributed by stellar captures.


  The total angular momentum of the accretion disk within $r_{1c}$
might, in fact, be self-regulated by the presence of an AGN wind.
If random fluctuations
in the number of plunging stars coming in with different
angular momenta cause the disk to lose a lot of angular momentum at some
point, the natural response is that as the disk shrinks the accretion
rate is increased and a strong AGN wind results when the luminosity
becomes too close to the Eddington value. Some of the ejected material
might then fall back into the disk at larger radius, mixing with gas of
high specific angular momentum and dragging some of that gas to smaller
radius.

  Another mechanism that may cause a mixing of material of different
specific angular momentum is the lifting of gas from the disk when
stars go through it. This was described by Zurek \etal (1994, 1996), who
suggested that the resulting star tails and debris might be the origin
of the broad emission lines in AGN.


\section{Discussion}

  Several models have been suggested that can explain the
$M_{BH}-\sigma_*$ relation. These fall broadly into two
distinct classes: models in which the growth of the black hole is
limited by either radiative or mechanical feedback from the active
nucleus (e.g., Silk \& Rees 1998; Haehnelt, Natarajan, \& Rees 1998;
Blandford \& Begelman 1999, 2004; King 2003; Wyithe \& Loeb 2003;
Norman \etal 2004) and models in which the matter available in the
bulge in stars or gas determine the feeding of the central black hole
(e.g., Zhao, Haehnelt, \& Rees 2002; Merritt \& Poon 2004).
The first class of models face the difficulty of explaining how enough
material accretes from the large distances required down to the central
engine via an accretion disk, as we describe in more detail below.
Moreover, it is not clear how an outflow can be sustained over the
time required to entirely expel all the matter that could potentially
be fed to the black hole at a future time, out to a large distance from
the host galaxy, without at the same affecting the accretion disk much
closer to the black hole, which is energizing the outflow. The second
class of models (to which our model
belongs), when using stars to grow the black holes, have been faced up
to now by the key problem that they do not account for the high-redshift
quasars and that the black holes grow only on very long timescales. 

  The model introduced in this paper proposes that the principal
mechanism by which accretion disks around quasars acquire their mass is
from stars that plunge through the disk. The increased cross section for
capturing a star by the disk, compared to a direct capture by the black
hole, allows black holes to be fed at a rapid rate, thereby avoiding the
timescale problems encountered by previous models in this class.
Furthermore, because the stellar matter is dissolved into a thin disk
before falling into the black hole, and then it accretes at high
radiative efficiency, quasars are naturally accounted for as the main
mechanism by which black holes grew. Accretion disks around quasars are
predicted to be much smaller than previously believed (although a small
fraction of the accreted matter may be carried in as gas from large
radius), thereby avoiding the disk gravitational instability problem.
The predicted $M_{BH}-\sigma_*$ relation is consistent with
observations, and is related by our model to fundamental parameters of
the accretion disk and properties of the stellar population.

  As pointed out previously by Ostriker (1983), Syer \etal (1991), and
others, it is inevitable that some stars will be captured through this
process because the dense nuclear regions of galaxies must always
contain stars. The question is whether or not star captures will occur
at the high rate necessary to feed the disk and account for the final
black hole mass. We have shown that the answer to this question is
affirmative provided that two crucial assumptions are satisfied: first,
that nuclear starbursts produce a roughly isothermal initial density
profile of stars with a velocity dispersion $\sigma_c$ similar to that
of the whole spheroidal component (with any core radius in the stellar
distribution being smaller than the zone of influence of the final black
hole, $r_c$); and second, that the rate at
which stars are brought into a nearly radial orbit due to dynamical
relaxation and the triaxiality of the potential (which can result in a
substantial fraction of stochastic orbits) is sufficient to keep the
loss-cylinder full. These assumptions may turn out to be wrong, and if
this is the case other explanations will need to be found for how enough
gas is transported to the inner accretion disk in quasars without
turning to stars along the way and for the $M_{BH}-\sigma_*$ relation.
But if the assumptions are correct, our model provides a unified
solution of these apparently unrelated problems.

  The detailed predicted form of the $M_{BH}-\sigma_*$ correlation is
still subject to theoretical uncertainties that will need to be
carefully analyzed as the model we have introduced here is developed
further. The surface density profile of the disk, which is the main
property that determines the form of the $M_{BH}-\sigma_c$ correlation,
is affected by the viscosity mechanism that causes the disk to accrete
to the black hole, by the accretion rate of the quasar disk, and by
the heating and addition of angular momentum to the disk associated with
the plunging stars and their transformation inside the disk.
Nevertheless, the simple analysis presented in this paper based
on a disk with no heat source except viscous dissipation and no
redistribution of angular momentum (plus the argument we give in
\S 5 that the effects of the plunging stars are likely to be a
truncation of the disk around radius $r_{1c}$, maintaining the basic
scalings of the disk surface density profile with mass, radius and
accretion rate) suggests that it is plausible that this model can
account for the observed correlation.

  This plunging star model has the added benefit of providing a simple
way to deliver the large mass of the black hole to a very small disk
around it. The difficulty of explaining how nuclear black holes have
grown so big in the standard (but hypothetical) scenario whereby gas
reaches the center by accretion through a disk extending to large radius
should not be overlooked (e.g., Begelman 2003). In fact, taking as a
typical example a black hole with $M_{BH}=10^8 \msun$ in a galaxy with
$\sigma_*=200 \kms$ and stellar mass $M_b\sim 10^{11} \msun$ within 5
kpc of the center, the radius initially containing a mass of $10^8
\msun$ is 5 pc for a singular isothermal profile (and even larger for
shallower profiles). It is very difficult to see how all this mass
could form a gaseous disk and then accrete inwards by three orders of
magnitude in radius over the lifetime of the quasar, without turning
into stars. Observationally, we see that star formation is a universal
phenomenon taking place in every galaxy that contains cold gas above
some critical surface density threshold (e.g., Kennicutt 1989; Martin \&
Kennicutt 2001), and that cold gas does not migrate very much in radius
before it turns to stars. And yet, in the standard quasar model one is
forced to assume that the same thing does not happen in the innermost
parts of galaxies, and that cold gas is efficiently transported to the
nucleus. Our model solves this problem by proposing that all this gas
does indeed turn into stars, and then the stars are captured by the disk
at small radius. In this way, the disk gravitational instability problem
is likely also solved. The disk can still extend to radius much larger
than $r_{1c}$ from additional gaseous material left over from star
formation and expelled by evolved stars (which could be a source of
angular momentum for the disk), but with a much lower accretion rate
than in the inner disk supplied by plunging stars. The low surface
density of this outer disk, plus the additional heating source provided
by plunging stars, can help prevent star formation at large radius (see
Sirko \& Goodman 2003 for models of marginally self-gravitating disks).

  Norman \& Scoville (1988) also suggested a model in which quasars are
fueled by stars after a nuclear starburst, but in their case all the
matter is expelled by evolved stars. Their model still does not account
for why the gas present in this nuclear region stops forming stars at
some point and starts accreting into the black hole instead. Our
model brings the stellar matter in the central cluster close to the
black hole directly, and provides a natural black hole mass at which
this process should stop.


  What other observational predictions can our model make? Clearly the
structure of the accretion disk should be modified by the capture of
stars, probably showing some characteristic feature in the surface
density and temperature profiles at the capture radius. A clear test
of this scenario will require much more advanced theoretical modeling
than we have done here to obtain predictions for the disk profiles, and
observations that can resolve the continuum emission of the disk. The
addition of angular momentum to the central parts of the disk by
plunging stars coming in along random orbits may generate warped inner
disks with a characteristic variability timescale that could have
observable consequences. The star tails generated by the plunging stars
may give rise to the broad emission line region, as discussed by Zurek
\etal (1994, 1996). Hydrodynamical simulations may be necessary here to
make any predictions that could be confronted with observational tests,
which could perhaps suggest some diagnostic for the rate at which
stellar collisions with the disk occur, testing if the rate is as high
as required by our model.

  The clearest prediction of our model is that the quasar phenomenon
must take place in the context of nuclear starbursts. Only a very
compact starburst can provide the high density of stars in the nucleus
that is necessary to feed the accretion disk. Quasars may always follow
an initial stage of growth within a highly obscured compact starburst
region, which
becomes highly ionized and transparent only after the quasar has
reached a high luminosity and cleared the surrounding dust. The light
contribution from the starburst around the quasar will be difficult to
discern, because the total mass of the starburst that forms the plunging
stars within radius $r_c$ is comparable to the total mass that will be
accreted by the black hole, and the radiative efficiency of gas
accretion to the black hole is much higher than the efficiency of
nuclear burning in stars.
The most straightforward observational test of our model would be to
image the central few parsecs around a luminous quasar and see if the
population of plunging stars is there, but the high resolution required
and the small amount of light coming from the stars compared to the
quasar may make this a difficult challenge. In our model, the size of
the nuclear starburst from which stars are captured is $r_c \sim 5\,
{\rm pc} (M_{BH}/10^8 \msun)^{8/15}$, much smaller than the highest
resolution images available from nearby Seyfert galaxies (Pogge \&
Martini 2002).


  The high stellar density of the required nuclear starbursts does not
need to be preserved at the present time, long after the quasar is
dead. Several dynamical phenomena can reduce the central stellar
density: if massive stars are dominant in nuclear starbursts, only the
light from the small fraction of mass that formed low-mass stars would
be visible at present. Moreover, a large fraction of the initial stellar
mass would have been lost in stellar winds and supernovae, resulting
in adiabatic expansion of the remaining stellar population. Mergers of
galaxies harboring nuclear black holes from old quasars will lead to
the merger of black holes, which eject the stars near the center and
create a wide core in the stellar profile (Milosavljevi\'c \& Merritt
2001). Diverse merger histories in different galaxies may introduce the
large variability observed in the slope of the innermost stellar
profiles in elliptical galaxies (Faber \etal 1997; Balcells, Graham,
\& Peletier 2004). Even though our
model predicts that $M_{BH}$ is correlated with the velocity dispersion
of the nuclear starburst at the small radius, $r_c$, from which most of
the captured stars come, the correlation with the velocity
dispersion can remain a tight one out to much larger radius if nuclear
starbursts tend to form with a universal density profile close to
isothermal. Then, after the inner part of the stellar distribution is
altered by repeated galaxy and black hole mergers, the remaining present
correlation would be tightest when expressed in terms of the velocity
dispersion at large radius.

  The model also predicts that the $M_{BH}-\sigma_*$ relation should be
independent of redshift, since it is imprinted at the time the quasars
formed, except for the effects of galaxy mergers and the passive
evolution of the stellar population in changing the velocity dispersion.
The small scatter of the relation is easier to understand if galaxy
mergers shift the black hole mass and velocity dispersion of galaxies in
a direction approximately parallel to the $M_{BH}-\sigma_*$ relation (as
seems to be implied by the Faber-Jackson relation). This suggests that
the $M_{BH}-\sigma_*$ relation may remain unmodified by mergers and
should then be close to constant with redshift.

  Another prediction of our model is that not all active galactic nuclei
should lie on the same $M_{BH}-\sigma_*$ relation as the inactive
black holes, but they should deviate from it in a way that depends on
the $L/L_{Edd}$ ratio. When a black hole becomes active after a nuclear
starburst has occurred, its mass may initially be small and the supply
of fuel from plunging stars will be more than sufficient to maintain an
Eddington luminosity, but as the black hole grows in mass and approaches
the final $M_{BH}-\sigma_*$ relation, the ratio $L/L_{Edd}$ has to
decrease if the quasar is to be sustained (as discussed in \S 2.5).
Hence, active nuclei with $L\simeq L_{Edd}$ should lie below the
relation, and they should gradually come closer to the relation as
$L/L_{Edd}$ decreases towards a minimum value at which the nuclear
activity typically ceases, with $M_{BH} \propto (L/L_{Edd})^{-5/7}
\sigma_*^{30/7}$ if the radiative efficiency $\epsilon$ is constant
(see eq. [\ref{msir}]). Recent data on Narrow line Seyfert I
galaxies suggest a correlation that is roughly along these lines
(Mathur \etal 2001; Grupe \& Mathur 2004; Mathur \& Grupe 2004).

  Finally, the $M_{BH}-\sigma_*$ relation should break down at a black
hole mass where the mass of the accretion disk becomes comparable to
that of a star. When that happens, the timescale for the disk to
accrete onto the black hole should be comparable to the mean time to
capture one star, so the disk can disappear when, by chance, no star is
captured over a long enough period of time. Approximating the mass of
the disk as $\pi r_{1c}^2 \Sigma(r_{1c})$, and from equations
(\ref{sdpq}) and (\ref{r1cf}), we find that the mass of the disk is
$1\msun$ at $M_{BH}\sim 10^{4.5} \msun$. This shows that the mechanism
we have presented here cannot continue to operate below this mass.

  In summary, we have presented a novel mechanism for the formation of
nuclear black holes in the centers of galaxies through the capture of
stars by a gaseous accretion disk. Our model provides a physical
connection between the stellar populations of bulges and black hole
growth and reproduces the observed $M_{BH}-\sigma_*$ relation for
reasonable input assumptions as a natural consequence.

\acknowledgements

  We are grateful to Martin Haehnelt, Jordi Isern, David Merritt, Martin
Rees, Scott Tremaine, and David Weinberg for many insightful
discussions. JAK acknowledges the support of a university fellowship at
Ohio State. This work was supported in part by grant NSF-0098515.

\begin{figure*}
\centerline{
\epsfxsize=8.5truein
\epsfbox[42 473 573 700]{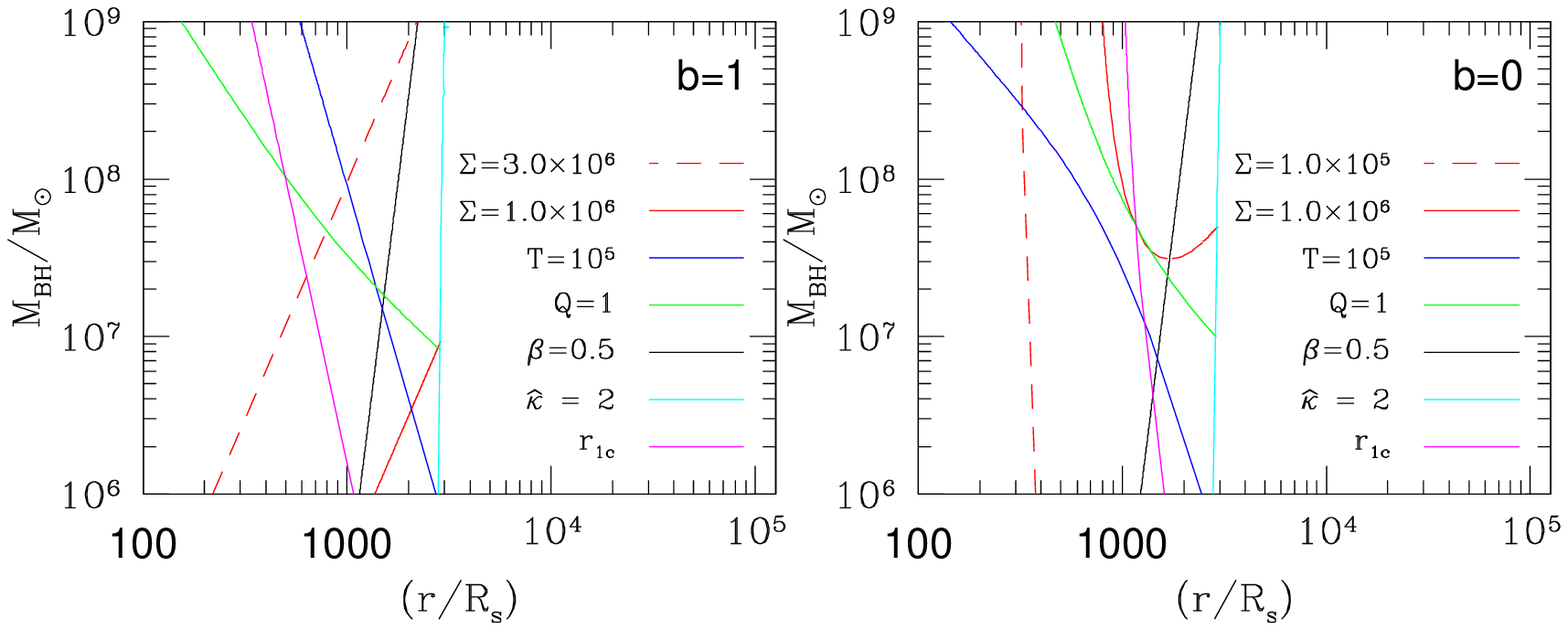}
}
\caption{Contours of accretion disk parameters $\Sigma$ (${\rm g\;}
{\rm cm}^{-2}$), T (K), {\rm Q}, $\beta$, $\hat\kappa$, and ${\rm
r}_{1c}/{\rm R_s}$ as indicated in each panel. {\it Left:} Contours
for the case of $b=1$, that is, the disk viscosity is proportional to
the gas pressure.  {\it Right:} Contours for the case of $b=0$, that
is, the disk viscosity is proportional to the total pressure.  All
contours are computed for $M_*/ M_{\odot}=1$, $\alpha_{-2}=1$,
$\dot m = 10$, and $\hat\kappa = 1$. We truncate curves to the right of
the $\hat\kappa=2$ contour, as they are no longer valid in this region.}
\label{fig:contourplot}
\end{figure*}

\end{document}